\documentclass{article}

\usepackage{arxiv}

\usepackage[utf8]{inputenc} 
\usepackage[T1]{fontenc}    
\usepackage{hyperref}       
\usepackage{url}            
\usepackage{booktabs}       
\usepackage{amsfonts}       
\usepackage{nicefrac}       
\usepackage{microtype}      
\usepackage{lipsum}
\usepackage{graphicx}
\graphicspath{ {./images/} }

\usepackage{comment}
\usepackage{tabularx} 
\usepackage{float}
\usepackage{subcaption} 

\usepackage{amsmath}
\usepackage{multirow}

\title{A County-Level Similarity Network of Electric Vehicle Adoption: Integrating Predictive Modeling and Graph Theory}

\author{
 Fahad S. Alrasheedi \\
  Department of Computer Science\\
  University of Nebraska at Omaha\\
  Omaha, NE 68182 \\
  \texttt{falrasheedi@unomaha.edu} \\
   \And
 Hesham H. Ali \\
  Department of Computer Science\\
  University of Nebraska at Omaha\\
  Omaha, NE 68182 \\
  \texttt{hali@unomaha.edu} \\
  }

\begin{document}
\maketitle
\begin{abstract}
Electric vehicle (EV) adoption is essential for reducing carbon dioxide (CO$_2$) emissions from internal combustion engine vehicles (ICEVs), which account for nearly half of transportation-related emissions in the United States. Yet regional EV adoption varies widely, and prior studies often overlook county-level heterogeneity by relying on broad state-level analyses or limited city samples. Such approaches risk masking local patterns and may lead to inaccurate or non-transferable policy recommendations. This study introduces a graph-theoretic framework that complements predictive modeling to better capture how county-level characteristics relate to EV adoption. Feature importances from multiple predictive models are averaged and used as weights within a weighted Gower similarity metric to construct a county similarity network. A mutual $k$-nearest-neighbors procedure and modularity-based community detection identify 27 clusters of counties with similar weighted feature profiles. EV adoption rates are then analyzed across clusters, and standardized effect sizes (Cohen’s $d$) highlight the most distinguishing features for each cluster. Findings reveal consistent global trends, such as declining median income, educational attainment, and charging-station availability across lower adoption tiers; while also uncovering important local variations that general trend or prediction analyses fail to capture. In particular, some low-adoption groups are rural but not economically disadvantaged, whereas others are urbanized yet experience high poverty rates, demonstrating that different mechanisms can lead to the same adoption outcome. By exposing both global structural patterns and localized deviations, this framework provides policymakers with actionable, cluster-specific insights for designing more effective and context-sensitive EV adoption strategies.
\end{abstract}


\section{Introduction}
\label{intro}

In the United States, nearly one-third of total CO\textsubscript{2} emissions originate from the transportation sector, and more than half of these emissions are attributed to light-duty vehicles~\cite{Us_Epa1, Us_Epa2}. To mitigate these emissions, U.S. authorities have implemented extensive plans and policies, including subsidies to expand the charging infrastructure network and financial incentives to promote the adoption of EVs over ICEVs~\cite{usa5}.

Despite these national efforts, EV adoption varies substantially across regions, indicating that policy incentives alone cannot fully explain why some areas adopt EVs rapidly while others lag behind. This variation has motivated a growing body of research seeking to identify the socio-demographic, economic, environmental, political, and infrastructure-related factors that shape EV adoption~\cite{usa1,usa2,usa3,usa4}. Most of these studies rely on survey-based or prediction-based analyses conducted on broad state-level populations or limited sets of cities, which restricts the generalizability of their findings. To address this gap, Kamis et~al.~\cite{usa3} incorporated county-level features across the entire United States; however, their target variable was assigned at the state level, meaning that all counties within the same state share the same EV adoption value. This approach masks important within-state variation and reduces the ability to detect local adoption patterns. Another relevant study by Alrasheedi et~al.~\cite{alrasheedi2025graph} employed graph-theoretic analysis at the county level to investigate the relationship between EV adoption and charging infrastructure, but it focused narrowly on this interaction and did not consider other influential determinants such as socioeconomic, political, or environmental features.

Consequently, existing analyses may be insufficient to reveal how contextual features jointly shape EV adoption across groups of similar counties. Generalizing feature importance across all regions without accounting for local differences can lead to misleading interpretations and ineffective policy design. Therefore, further investigation is needed to examine how influential features vary not only between broad adoption tiers (\textit{i.e.}, high, medium, low adoption) but also within clusters of counties that share similar characteristics. This makes the transition from ICEVs to EVs a continuing management challenge for policymakers worldwide.

To address this challenge, this study proposes a graph-theoretic framework that complements prediction-based analyses by uncovering structural similarities among counties based on key features influencing EV adoption. Our approach integrates socioeconomic, political, environmental, and infrastructure-related variables, weighted by their average importance across several predictive models. These weighted features are used to compute pairwise similarities between counties. Because the dataset includes both numerical and categorical variables, we employ the weighted Gower distance to measure dissimilarity between counties, as explained in Section~\ref{similarity_method}. The resulting dissimilarity values, ranging from 0 to 1, are converted to similarity scores, which are used to construct a mutual $k$-nearest neighbor ($kNN$) network. The value of $k$ is chosen automatically using a modularity-based optimization algorithm that maximizes intra-cluster similarity while minimizing inter-cluster similarity. From this process, we identify 772 counties (out of 1888) forming 27 clusters that exhibit strong feature-based similarity patterns.

Importantly, the target variable, EV adoption rate, is excluded from the similarity computation to ensure that clustering is driven solely by contextual features. EV adoption is then introduced as an external parameter to assess whether counties with similar characteristics also exhibit similar EV adoption patterns. The results reveal clear differences among clusters, which naturally fall into three tiers: High Adoption Tier (HAT), Medium Adoption Tier (MAT), and Low Adoption Tier (LAT). A global  Permutational Multivariate Analysis of Variance test (\textit{i.e.}, PERMANOVA) confirms strong statistical evidence (p-value < 0.001) that the clusters differ significantly in their combined features. Subsequent pairwise tests reinforce these differences: nearly all of the 351 pairwise comparisons show significant distinctions between clusters. For each cluster, we compute standardized mean differences (Cohen’s $d$) for all features and identify the four most distinguishing features.

A key contribution of this study lies in revealing how features behave not only across tiers but also within them. Several features (such as median income, educational attainment, and charging-station availability) exhibit consistent negative trends in their effect sizes as EV adoption decreases, indicating global structural patterns that shape adoption nationwide. Although minor local fluctuations exist, these features agree strongly within each cluster and across clusters, reinforcing their role as broad socioeconomic determinants.

In contrast, urbanicity and poverty demonstrate a positive trend across the adoption tiers but exhibit substantial disagreement within the LAT. Some clusters in the LAT consist of counties that are highly rural yet not economically disadvantaged, suggesting that rurality, rather than income, might be the primary barrier to EV adoption in these areas. Other clusters in the same tier, however, are relatively urbanized but experience high poverty rates, indicating that economic constraints dominate despite favorable urban conditions. These feature-specific deviations are further confirmed by the cluster-level significance results, which show that poverty strongly characterizes the urbanized–high-poverty clusters, while urbanicity is a key distinguishing feature in the rural–low-poverty clusters.

These findings highlight the interpretive power of the graph-theoretic approach: clusters with similar contextual profiles may exhibit different mechanisms leading to low EV adoption. Unlike general trend or predictive analyses, which tend to smooth out local deviations, the graph framework reveals the nuanced and diverse pathways that drive adoption outcomes within and across tiers. This level of granularity enables policymakers to design targeted, cluster-specific strategies that address the underlying socioeconomic or infrastructural barriers unique to each region, rather than relying on one-size-fits-all interventions.

The remainder of this paper is organized as follows. Section~\ref{lit} reviews the related literature on EV adoption. Section~\ref{metho} describes the prediction analysis and presents the graph-theoretic methodology used to construct the similarity network and identify clusters. Section~\ref{result} reports the empirical findings from the predictive models, the similarity network, and the cluster analysis. Finally, Section~\ref{conclusion} summarizes the key insights and concludes the paper.

\section{Literature Review}
\label{lit}

There exists an extensive body of research investigating the determinants of EV adoption, often through survey-based approaches grounded in behavioral theories. For example, studies have applied the extended Unified Theory of Acceptance and Use of Technology (UTAUT2) ~\cite{khazaei2021moderating,huang2021influence}
, the Value-Belief-Norm (VBN) model~\cite{higueras2024factors}
, the Diffusion of Innovation theory~\cite{xia2022economic}
, the Theory of Planned Behavior~\cite{yeugin2022analysis,buhmann2024predicting}
, the Perceived Risk-Benefit and Norm models~\cite{choo2024predicting} 
, the Task-Technology Fit and Expectation-Confirmation frameworks~\cite{cruz2023pragmatic} 
, the Technology Acceptance Model (TAM)~\cite{shanmugavel2022exploring}
, the Stimulus-Organism-Response theory~\cite{upadhyay2023examining} 
, and Maslow’s Hierarchy of Needs~\cite{cui2021predicting}. 

Most of these studies employ factor analysis and structural equation modeling to validate relationships between psychological factors and the intention to adopt EVs. However, they focus primarily on intention rather than actual adoption trends, and they rely on survey data rather than historical adoption data. Furthermore, these studies often focus on countries or large regions rather than U.S. counties, limiting their ability to capture localized patterns.

Due to the limited availability of fine-grained data, many prediction-based studies have been conducted at the national or state level. For example, Ruoso et al.~\cite{ruoso2022influence}
 used socioeconomic factors for 28 countries, both developed and developing, to model adoption dynamics using nonlinear regression and S-curves. Similarly, Brückmanna et al.~\cite{bruckmann2021battery} 
 employed a generalized linear mixed-effects logistic model to study spatial and individual effects on EV adoption in Switzerland. These studies used revealed preference (RP) survey data, focusing on previous adopters, and largely excluded the U.S. context.

Other works, such as Ding et al.~\cite{ding2021forecasting} 
, modified the traditional grey prediction model to improve forecasting accuracy under data uncertainty by optimizing initial conditions and applying Simpson’s formula for smoothing. However, such models are univariate, focusing solely on forecasting EV sales or stock over time without considering multi-factorial drivers.

In the U.S. context, Archsmith et al. 
analyzed survey data from 2017–2018 to forecast future vehicle purchase patterns under different assumptions about consumer preferences, technology improvements, and cost declines~\cite{archsmith2022future}. However, the study operated at the state level, missing intra-state county-level heterogeneity. Another study by Forsythe et al. 
examined how consumer preferences for EVs evolved between 2013 and 2021 as technology advanced~\cite{forsythe2023technology}, while Soltani-Sobh et al. 
conducted a panel analysis (2003–2011) linking EV market share to electricity price, gas price, income, and incentives ~\cite{soltani2015investigating}. Yet, these datasets are outdated and aggregated at the state level, limiting insight into recent, local adoption patterns.

More recently, Kamis et al. 
developed models to predict the number of EVs, chargers, and EV-related jobs for 2022 using county-level features but state-level targets, leading to inflated accuracies due to target homogeneity within states~\cite{usa3}. Our study addresses this limitation by using both features and targets at the county level, incorporating additional predictors such as average gas price and political affiliation, and relying on more recent data.

Finally, Alrasheedi et al. 
conducted a graph-theoretic analysis exploring correlations between EV adoption and charging infrastructure expansion across 137 U.S. counties~\cite{alrasheedi2025graph}. Although valuable, this work focused solely on the interaction between EV and charging infrastructure growth, excluding other important factors such as socioeconomic, political, and environmental attributes. The present study builds upon and extends this approach by integrating a broader set of determinants to construct a comprehensive similarity network of EV adoption across U.S. counties.

\section{Methodology}
\label{metho}

\subsection{Target Variable}
\label{target}

One of the main challenges in EV related research is the limited availability of county-level data. Focusing on the year 2023 allows the collection of data for 28 states, covering approximately 1,888 counties. Among these, data for 20 states are obtained directly from official sources
~\cite{s1_Maryland_2025,
s2_Pennsylvania_PennDOT_2025,
s3_NewYork_NYSERDA_2025,
s4_Oregon_2025,
s5_Utah_2025,
s6_Texas_DFW_2025,
s7_Illinois_2025,
s8_NorthCarolina_2025,
s9_Connecticut_2025,
s10_Massachusetts_2025,
s11_Vermont_2025,
s12_Maine_2025,
s13_Iowa_2025,
s14_Hawaii_2025,
s15_Ohio_2025,
s16_Pennsylvania_PennShare_2025,
s17_Michigan_SEMCOG_2025,
s18_Louisiana_2025,
s19_Washington_2025,
s20_California_CEC_2025}, while data for 8 states are collected from the Atlas EV Hub~\cite{Nigro_undatedj}, which compiles information from official state databases.

\begin{table}[H]
\centering
\caption{Descriptive statistics of EV counts before and after normalization by population and scaling to EVs per 10,000 people.}
\label{tab:ev_norm_stats}
\resizebox{\columnwidth}{!}{%
\begin{tabularx}{\textwidth}{lrr}
\toprule
\textbf{Statistic} & \textbf{Raw Data} & \textbf{After Transformation} \\
\midrule
Count & 1888 & 1888 \\
Mean  & 611.99 & 16.53 \\
Standard Deviation & 3741.84 & 22.98 \\
Minimum & 0.00 & 0.00 \\
25th Percentile & 10.00 & 3.50 \\
Median (50th Percentile) & 30.00 & 8.76 \\
75th Percentile & 160.00 & 20.17 \\
Maximum & 123{,}380.00 & 216.24 \\
\bottomrule
\end{tabularx}%
}
\end{table}

This study focuses primarily on vehicles that can be recharged through external chargers, specifically Battery Electric Vehicles (BEVs) and Plug-in Hybrid Electric Vehicles (PHEVs). Consequently, the target variable is the number of EV registrations in 2023 (EV counts). However, since there is a large variation in EV counts across counties due to population differences, the EV count for each county was normalized by its population and then scaled per 10,000 people. Table \ref{tab:ev_norm_stats} presents the descriptive statistics of EV counts before and after normalization and scaling. 


\begin{table}[ht]
\centering
\caption{Summary of datasets and corresponding categories used in this study. 
The data sources include major U.S. federal agencies and research institutions such as the EPA, NOAA, BEA, USDA, EIA, Census Bureau, DOE, CDC, and MEDSL. 
All datasets were harmonized to align with the 2020 coverage year used in Paper~2.}
\begin{tabularx}{\linewidth}{l l l c c}
\toprule
\textbf{Feature} & \textbf{Source} & \textbf{Category} & \textbf{Time} & \textbf{Kamis et al.} \\
\textbf{Name} &  &  & \textbf{Coverage} & \textbf{Coverage} \\
\midrule
Air Quality (AQI) & EPA~\cite{aqi} & Environmental / Health & 2022 & 2020 \\
Mean Temperature & NOAA~\cite{temp} & Environmental / Climate & 2022 & 2021 \\
GDP & BEA~\cite{gdp} & Economic & 2022–2023 & 2020 \\
Education Level & USDA (ACS)~\cite{education} & Socioeconomic / Education & 2018–2022 & 2013-2017 \\
Urbanicity & USDA ERS~\cite{Urbanicity} & Demographic / Spatial & 2023 & 2013 \\
Solar Generation & EIA~\cite{sun} & Energy / Renewable & 2022 & 2019 \\
Median Income & Census (ACS)~\cite{income} & Finance & 2018–2022 & 2019 \\
Poverty Rate & Census (ACS)~\cite{poverty} & Economic / Finance & 2018–2022 & 2019 \\
Gas Price & Gas Buddy~\cite{gas} & Energy / Economic & 2022 & - \\
Political Affiliation & MIT Election Lab~\cite{election} & Political / Demographic & 2000-2020 & - \\
Smoking Ban & CDC~\cite{somking_ban} & Health Policy / Policy & 2023 & 2010 \\
EV Charging Stations & DOE AFDC~\cite{charging} & Energy / Infrastructure & 2022 & 2021 \\
EV Registrations & DOE AFDC~\cite{doe_afdc_ev_registrations_2022} & Transportation / Adoption & 2022-2023 & 2021 \\
\bottomrule
\end{tabularx}
\label{tab:features_summary}
\end{table}

\subsection{Feature Variables}
\label{features}

The features used in this study are similar to those adopted in previous research in Section~\ref{lit}, and most have been identified as significant determinants of EV diffusion~\cite{usa3,soltani2015investigating}. The selected features span multiple domains, including health, environment, education, economy, finance, energy, political, and geography. Unlike earlier studies that primarily relied on data from 2010–2020, the present study incorporates more recent datasets, generally corresponding to one year prior to the target variable year. Table \ref{tab:features_summary} summarizes all features, their data sources, and years of coverage, alongside a comparison with Kamis et al.~\cite{usa3}. All numerical features were scaled to the range [0, 1] to prevent large values from disproportionately influencing the modeling and similarity analyses. Table \ref{tab:feature_stats_before_after} presents the descriptive statistics of these features before and after scaling. Additionally, average gasoline price and political affiliation were introduced as new variables to explore their potential influence on EV adoption. Finally, Table~\ref{tab:categorical_summary} reports the descriptive statistics for the categorical variables.

\begin{table*}[ht]
\centering
\caption{Comparison of numerical feature statistics before and after normalization and scaling ($n = 1888$). 
Variables already bounded between 0 and 1 (\textit{Poverty}, \textit{High School Graduate}, \textit{College Completion}) were not scaled.}
\begin{tabularx}{\textwidth}{l c c c c c}
\toprule
\textbf{Feature} & \textbf{Stage} & \textbf{Mean} & \textbf{Std.} & \textbf{Min} & \textbf{Max} \\
\midrule
Sun Energy Generation & Before & 59211.68 & 360697.19 & 0.00 & 9574576.09 \\
                      & After  & 0.006 & 0.038 & 0.000 & 1.000 \\
Charging Units per 10k & Before & 10.21 & 30.46 & 0.00 & 833.57 \\
                      & After  & 0.012 & 0.037 & 0.000 & 1.000 \\
Median Income ($\times 10^4$ USD) & Before & 6.50 & 1.75 & 2.74 & 17.05 \\
                      & After  & 0.263 & 0.122 & 0.000 & 1.000 \\
College Completion Rate & Before & 0.24 & 0.11 & 0.00 & 0.77 \\
                      & After  & 0.24 & 0.11 & 0.00 & 0.77 \\
High School Graduate Rate & Before & 0.89 & 0.06 & 0.57 & 1.00 \\
                      & After  & 0.89 & 0.06 & 0.57 & 1.00 \\
Avg. Gas Price (USD/gal) & Before & 2.96 & 0.38 & 2.34 & 5.48 \\
                      & After  & 0.197 & 0.120 & 0.000 & 1.000 \\
Poverty Rate & Before & 0.14 & 0.06 & 0.00 & 0.48 \\
                      & After  & 0.14 & 0.06 & 0.00 & 0.48 \\
Median AQI & Before & 40.28 & 6.56 & 2.00 & 80.00 \\
                      & After  & 0.491 & 0.084 & 0.000 & 1.000 \\
Average Temperature (°F) & Before & 55.62 & 8.20 & 36.10 & 77.40 \\
                      & After  & 0.473 & 0.199 & 0.000 & 1.000 \\
Urbanicity Index & Before & 4.77 & 2.85 & 1.00 & 9.00 \\
                      & After  & 0.471 & 0.356 & 0.000 & 1.000 \\
GDP ($\times 10^6$ USD) & Before & 10.72 & 43.44 & 0.03 & 919.93 \\
                      & After  & 0.012 & 0.047 & 0.000 & 1.000 \\
Total EVs (2022) & Before & 113115.84 & 212856.10 & 5000.00 & 1264700.00 \\
                      & After  & 0.086 & 0.169 & 0.000 & 1.000 \\
Stations per 10k & Before & 2.03 & 3.73 & 0.00 & 45.23 \\
                      & After  & 0.045 & 0.083 & 0.000 & 1.000 \\
\bottomrule
\end{tabularx}
\label{tab:feature_stats_before_after}
\end{table*}

\begin{table}[h!]
\centering
\caption{Summary of Categorical Variables}
\label{tab:categorical_summary}
\begin{tabular}{lcc}
\toprule
\textbf{Variable} & \textbf{Category} & \textbf{Count (\%)} \\
\midrule
\multirow{3}{*}{Smoking Ban} 
    & None          & 1383 (73.25\%) \\
    & Partial       & 301 (15.94\%) \\
    & Comprehensive & 204 (10.81\%) \\
\midrule
\multirow{2}{*}{Political Affiliation} 
    & Republican & 1453 (76.96\%) \\
    & Democrat   & 435  (23.04\%) \\
\bottomrule
\end{tabular}
\end{table}

 Obtaining average gasoline price data at the county level for 2022 was challenging because most publicly available datasets report prices only at the state level~\cite{EIA_GasDiesel_2025}. To overcome this limitation, gasoline prices from 20 randomly selected stations per county were collected from Gas Buddy~\cite{gas}, averaged within each county, and then adjusted to reflect 2022 values using the national average decline ratio of 21.7\%, calculated from the U.S. average prices in April 2025 (\$3.299 per gallon) and April 2022 (\$4.213 per gallon)~\cite{eia_retail_gasoline_all_grades_2025}. 

For political affiliation, official county-level election data were obtained from MIT Election Lab~\cite{election}. These data include the winning party (Republican or Democrat) in the 2002, 2006, 2010, 2014, 2018, and 2022 elections. Each county’s political affiliation was determined based on the most frequently winning party across these six election cycles.

\subsection{Prediction Models}

This study employs both linear and nonlinear models to predict the target variable in Section \ref{target}. The linear models include Ridge Regression (RR), Elastic Net Regression (ENR), and Huber Regressor (HR), while the nonlinear models comprise Random Forest (RF), Gradient Boosting Tree (GBT), Histogram-Based Gradient Boosting Tree (HGBT), and Multi-Layer Perceptron (MLP).

RR, also known as L2-regularized linear regression, introduces a penalty on the squared L2 norm of the coefficients to control overfitting and mitigate multicollinearity. ENR, combines both L1 and L2 regularizations, allowing for simultaneous feature selection and regularization by shrinking some coefficients to zero. HR, on the other hand, is a robust linear model resistant to outliers; it employs a piecewise loss function that behaves quadratically for small residuals and linearly for large ones. 



RF is an ensemble of decision trees in which each tree is trained on a random subset of the data, and the final prediction is obtained by averaging the outputs of all trees. This approach reduces both variance and overfitting. GBT builds trees sequentially, where each new tree is trained to correct the residual errors of the previous ensemble, thereby improving prediction accuracy. HGBT follows the same principle as GBT but speeds up computation by grouping continuous features into discrete bins during training. Finally, MLP introduces nonlinearity through hidden layers and activation functions, enabling it to capture complex relationships between features and EV adoption.



\subsection{Model Training and Validation Procedure}

We trained all models using stratified 5-fold cross-validation. The target variable was first binned for stratification using quantile cuts with up to ten bins, selected empirically. The stratified 5-fold cross-validation procedure was repeated five times, with the data reshuffled before each repetition. This resulted in 25 total train/validation splits (5 folds × 5 repeats), ensuring that each fold received an approximately equal distribution of target bins. In each split, the model was trained on four folds and evaluated on the held-out fold. The validation metrics were then averaged across all 25 runs to obtain the reported performance. This approach helps stabilize performance estimates and preserve the marginal distribution of the (binned) target within each fold.

\subsection{Similarity Computation}
\label{similarity_method}




The features are weighted based on the average of their corresponding permutation importances across the top three predictive models. 
Because the dataset contains both numerical and categorical variables, we compute pairwise dissimilarities using a weighted Gower distance~\cite{gower1971general}.

In the first step, we compute the dissimilarity between each pair of counties using the weighted Gower distance, which combines a normalized absolute difference for numerical features and a weighted mismatch indicator for categorical features. The dissimilarity between nodes \(i\) and \(j\) is given by:

\[
D_{ij} =
\frac{
\sum_{f \in F_n} w_f \, \frac{|x_{if} - x_{jf}|}{R_f}
+
\sum_{f \in F_c} w_f \, \delta(x_{if}, x_{jf})
}{
\sum_{f \in F_n \cup F_c} w_f
},
\]

where \(w_f\) is the feature weight, \(R_f = \max(x_{\cdot f}) - \min(x_{\cdot f})\) is the range of numerical feature \(f\), and 
\(\delta(x_{if}, x_{jf}) = 0\) if the categorical values match and \(1\) otherwise.

We then convert dissimilarity to similarity as:

\[
S_{ij} = 1 - D_{ij}, 
\qquad S_{ii} = 0,
\]

where setting \(S_{ii}=0\) removes self-loops in the similarity network, and higher values of \(S_{ij}\) indicate greater similarity.

\subsection{Similarity Network Construction via Mutual \texorpdfstring{$k$}{k}-Nearest Neighbors}

After computing the pairwise similarity matrix using the weighted Gower distance, we construct the county-level similarity network using a mutual $k$-nearest neighbors ($k$NN) approach~\cite{maier2009optimal}. In this framework, an undirected edge is formed between two counties only if each county appears among the other's top $K$ most similar neighbors. The value of $K$ plays a critical role in determining the connectivity, sparsity, and community structure of the resulting network; the following section explains how the optimal $k$ is selected.

\subsubsection{Selection of the Optimal $K$}
\label{optimalK}
Rather than specifying $K$ manually, we determine it automatically by optimizing a balance between modularity, network connectivity, and density. For each candidate value of $K$, we compute two structural properties of the resulting graph:

\begin{itemize}
    \item \textbf{Isolate fraction}:
\[
\mathrm{IF}(K) = \frac{\text{number of nodes with degree } 0}{|V|},
\]
    which measures the proportion of isolated nodes produced by the choice of $K$~\cite{newman2010networks}.
    
    \item \textbf{Average degree}:
    \[
    AD(K) = \frac{2|E|}{|V|},
    \]
    where $|E|$ and $|V|$ denote the numbers of edges and nodes, respectively~\cite{newman2010networks}.
\end{itemize}

These quantities guide the application of sparsity and connectivity penalties to avoid undesirable graph structures.

\subsubsection{Penalty Functions for Network Disconnection and Sparsity Control}

To ensure that the similarity network remains neither excessively sparse nor unrealistically 
dense, we incorporate soft penalty terms into the $K$-selection procedure. Extremely sparse 
graphs—such as those with an average degree near 1 or below—tend to fragment and fail to 
reveal meaningful community structure, whereas overly dense graphs approach a nearly complete 
network and obscure local similarity patterns.

In mutual $k$-nearest-neighbor (kNN) graphs, prior work shows that stable, cluster-preserving 
connectivity typically arises when the average degree lies between 2 and 3, a range that 
balances interpretability, modularity, and robustness~\cite{luxburg2007spectral}. Rather 
than enforcing a hard threshold, we introduce a soft average-degree penalty that lightly 
discourages excessively sparse networks and more strongly penalizes networks that become 
too dense relative to $K$:

\[
P_{AD}(K) =
\begin{cases}
2 - AD(K), & \text{if } AD(K) < 2, \\[6pt]
AD(K) - K, & \text{if } AD(K) > K, \\[6pt]
0,         & \text{otherwise}.
\end{cases}
\]

where $AD(K)$ is the average degree of the graph constructed with parameter $K$, as defined 
in Section~\ref{optimalK}. This formulation guides the selection of $K$ toward networks that 
remain connected enough to support modularity-based clustering while retaining deliberate 
sparsity that preserves meaningful local structure.

Moreover, to discourage disconnected solutions, we impose an isolate-fraction penalty:
\[
P_{IF}(K) = \max\bigl(0,\, IF(K) - 0.05\bigr),
\]
where $IF(K)$ is the isolate fraction (Section~\ref{optimalK}). This penalty activates only 
when more than 5\% of nodes become isolated. Together, these penalty functions promote 
similarity networks with stable connectivity, interpretable sparsity levels, and strong 
modularity properties.

\subsubsection{Scoring Function and Modularity Maximization}

Each candidate $K$ is evaluated using the scoring function:
\[
\text{Score}(K) = \text{Modularity}(K) - 10\,P_{IF}(K) - P_{AD}(K).
\]
where modularity measures the degree to which the network exhibits well-defined community structure. We use the standard definition~\cite{newman2004finding}:
\[
Q(K) = \frac{1}{2m}
\sum_{i,j}
\left(A_{ij} - \frac{k_i k_j}{2m}\right)
\delta(c_i, c_j),
\]
with $A_{ij}$ denoting the adjacency matrix, $k_i$ and $k_j$ the degrees of nodes $i$ and $j$, 
$m = |E|$ the total number of edges, and $\delta(c_i,c_j)=1$ when nodes $i$ and $j$ belong to the same community.

The optimal value of $K$ is selected as the one that maximizes $\text{Score}(K)$, producing a similarity network that is well-connected, avoids excessive graph density, and yields strong community structure.

\subsubsection{Community Detection and Cluster Labeling}

Once the optimal $K$ is identified, the final mutual $k$NN graph is constructed. Modularity-based community detection is then applied to partition the graph into clusters. Each detected community is treated as a cluster, and each county is assigned a unique cluster label. These clusters serve as the foundation for the tiered EV adoption analysis presented in Section~\ref{result}.

\section{Results}
\label{result}

\subsection{Predictive Modeling Based Analysis}

This section presents the correlation and multicollinearity diagnostics for the numerical features. It then compares the performance of linear models and nonlinear models. Additionally, the feature importances for the top three models, along with their average values, are reported. Finally, this section provides a general comparison between our modeling results and findings from the existing literature.

\subsubsection{Correlation and Multicollinearity}

    \begin{figure}[h!]
        \centering
        \includegraphics[width=0.8\textwidth]{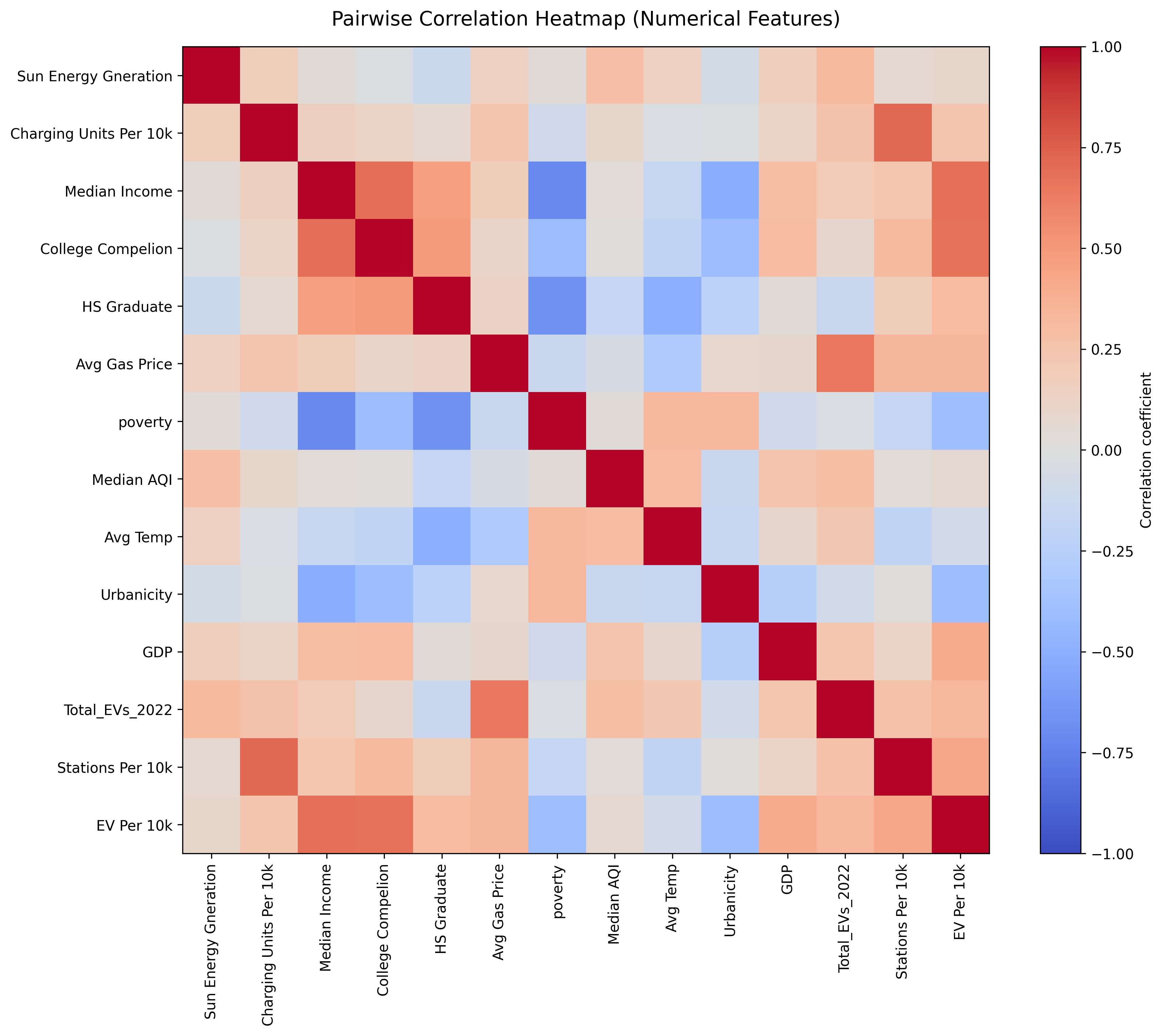}
        \caption{A descriptive caption for your figure.}
        \label{fig:r1}
    \end{figure}

Figure \ref{fig:r1} \ shows the correlation matrix for all numerical variables, including the target variable after the transformation as explained in Section~\ref{target}. There are no strong correlations among the features; all pairwise correlations are below 0.7.

\begin{table}[H]
\centering
\caption{Variance Inflation Factor (VIF) for the selected features used in the analysis.}
\label{tab:r1}
\scriptsize
\setlength{\tabcolsep}{6pt}
\renewcommand{\arraystretch}{1.1}
\begin{tabularx}{\textwidth}{c l c}
\toprule
\textbf{Group Number} & \textbf{Feature} & \textbf{VIF} \\
\midrule
Group 1 (singleton)  & High School Graduate       & 0.308 \\
Group 2 (singleton)  & Total\_EVs\_2022           & 0.336 \\
Group 3 (singleton)  & Avg Gas Price              & 0.338 \\
Group 4 (singleton)  & Stations Per 10k           & 0.426 \\
Group 5 (singleton)  & Avg Temperature            & 0.076 \\
Group 6 (singleton)  & Urbanicity                 & 0.400 \\
Group 7 (singleton)  & Median Income              & 0.682 \\
Group 8 (singleton)  & Poverty                    & 0.396 \\
Group 9 (singleton)  & Sun Energy Generation      & 0.092 \\
Group 10 (singleton) & College Completion         & 0.676 \\
Group 11 (singleton) & Charging Units Per 10k     & 0.236 \\
Group 12 (singleton) & GDP                        & 0.402 \\
Group 13 (singleton) & Median AQI                 & 0.066 \\
\bottomrule
\end{tabularx}
\end{table}

To further assess multicollinearity, we computed the Variance Inflation Factor (VIF) for all features. Table \ref{tab:r1} presents these results, showing that all VIF values are below 10, indicating no serious multicollinearity problems. Furthermore, both the linear and nonlinear predictive models used in this study are specifically designed to handle moderate correlations among features.

\subsubsection{Linear Models}

\begin{table}[H]
\centering
\caption{Linear Model performance comparison based on mean and standard deviation across repeated cross-validation runs.}
\label{tab:r2}
\scriptsize
\setlength{\tabcolsep}{5pt}
\renewcommand{\arraystretch}{1.1}
\begin{tabularx}{\textwidth}{l c c c c c c}
\toprule
\textbf{Model} & \textbf{R$^2$ (mean)} & \textbf{R$^2$ (std)} & \textbf{MAE (mean)} & \textbf{MAE (std)} & \textbf{RMSE (mean)} & \textbf{RMSE (std)} \\
\midrule

Ridge       & 0.647 & 0.042 & 8.73 & 0.37 & 13.55 & 1.10 \\
ElasticNet  & 0.523 & 0.037 & 9.67 & 0.51 & 15.81 & 1.48 \\
Huber       & 0.624 & 0.082 & 8.08 & 0.43 & 13.92 & 1.50 \\
\bottomrule
\end{tabularx}
\end{table}

Table \ref{tab:r2} compares the performance of the linear models (Ridge Regression, Elastic Net, and Huber Regressor) using three evaluation metrics: coefficient of determination ($R^2$), root mean square error (RMSE), and mean absolute error (MAE). All metrics are computed exclusively on the held-out validation folds, not on the training data. Each evaluation metric is shown with the mean and standard deviation across 25 runs of repeated stratified 5-fold cross-validation. 


The small standard deviations indicate that the models exhibit stable and consistent performance across different splits of the data. Among the linear models, Ridge Regression achieves the best overall performance, yielding the highest $R^2$ and the lowest error values, while the Huber Regressor performs comparably but with slightly higher variability.

\begin{table}[H]
\centering
\caption{Model coefficients in scaled and original feature units for the Ridge model.}
\label{tab:r4}
\scriptsize
\setlength{\tabcolsep}{6pt}
\renewcommand{\arraystretch}{1.1}
\begin{tabularx}{\textwidth}{l c r r}
\toprule
\textbf{Feature} & \textbf{Type} & \textbf{Coefficient (scaled)} & \textbf{Effect (original units)} \\
\midrule
College Completion              & numerical & 47.505224  & 61.607704  \\
Poverty                         & numerical & -7.054573  & -14.666471 \\
Avg Gas Price                   & numerical & 27.146938  & 8.665760   \\
HS Graduate                     & numerical & -1.627292  & -3.789371  \\
Political Affiliation\_DEMOCRAT & categorical& 3.768473   & 3.768473   \\
Political Affiliation\_REPUBLICAN & categorical& -3.768473 & -3.768473  \\
Median Income                   & numerical & 50.108093  & 3.501883   \\
Smoking Ban\_comprehensive      & categorical& -1.631573  & -1.631573  \\
Smoking Ban\_partial            & categorical& 1.118353   & 1.118353   \\
Stations Per 10k                & numerical & 35.483997  & 0.784577   \\
Urbanicity                      & numerical & -5.753687  & -0.719211  \\
Smoking Ban\_none               & categorical& 0.513221   & 0.513221   \\
Avg Temp                        & numerical & 8.610438   & 0.208485   \\
Median AQI                      & numerical & -2.673959  & -0.034282  \\
GDP                             & numerical & 28.679509  & 0.031177   \\
Charging Units Per 10k          & numerical & —          & -0.002266  \\
Total\_EVs\_2022                & numerical & —          & 0.000008   \\
Sun Energy Generation           & numerical & —          & 0.00000015 \\
\bottomrule
\end{tabularx}
\end{table}

Table \ref{tab:r4} presents the coefficients of the best-performing model, Ridge Regression. These coefficients were computed after scaling all features to the range [0, 1] as explained in Section~\ref{features}. To facilitate interpretation, we also converted these coefficients back to their effects in the original data scale in the same table .

The most influential positive features is College Completion, with a coefficient of 61.61. This indicates that an increase of one–percentage-point increase (equivalent to a 10\% increase in the raw data) in college completion corresponds to an increase of approximately 6.2 EVs per 10,000 people. In contrast, Poverty exhibits the largest negative effect (–14.67), implying that an increase of one scaled unit (approximately 10 percentage points) in poverty is associated with a decrease of about 1.47 EVs per 10,000 people.

Features positively associated with EV adoption include average gas price (8.7), political affiliation (Democrat) (3.8), median income (3.5), smoking ban (partial) (1.1), and number of charging stations per 10k (0.8). Negatively associated features include high-school graduation rate (–3.8), political affiliation (Republican) (–3.8), smoking ban (comprehensive) (–1.6), and urbanicity (–0.7). The latter implies that moving one unit toward rural areas on the urbanicity scale corresponds to a decrease of approximately 0.7 EVs per 10,000 people.

\subsubsection{Nonlinear Models}

\begin{table}[H]
\centering
\caption{Non-Linear Model performance comparison based on mean and standard deviation across repeated cross-validation runs.}
\label{tab:r3}
\scriptsize
\setlength{\tabcolsep}{5pt}
\renewcommand{\arraystretch}{1.1}
\begin{tabularx}{\textwidth}{l c c c c c c}
\toprule
\textbf{Model} & \textbf{R$^2$ (mean)} & \textbf{R$^2$ (std)} & \textbf{MAE (mean)} & \textbf{MAE (std)} & \textbf{RMSE (mean)} & \textbf{RMSE (std)} \\
\midrule
GBRT        & 0.765 & 0.037 & 6.42 & 0.32 & 11.02 & 0.81 \\
HGB         & 0.761 & 0.027 & 6.42 & 0.34 & 11.16 & 0.90 \\
RF          & 0.739 & 0.028 & 6.71 & 0.36 & 11.68 & 1.09 \\
MLP         & 0.606 & 0.090 & 7.81 & 0.53 & 14.18 & 1.48 \\
\bottomrule
\end{tabularx}
\end{table}

Table \ref{tab:r3} summarizes the results for the nonlinear models: Gradient Boosting Regression Tree (GBRT), Histogram-Based Gradient Boosting (HGB), Random Forest (RF), and Multi-Layer Perceptron (MLP)—using the same performance metrics. The GBRT model outperforms all others, achieving the highest $R^2$ and the lowest error values. HGB ranks second and performs very similarly to GBRT.

In general, the nonlinear models demonstrate substantially higher predictive performance than the linear models. This result suggests that EV adoption is governed by nonlinear relationships among socio-political, economic, environmental, and infrastructural factors, a finding that aligns with previous studies in Section \ref{lit}.

\subsubsection{Feature Importance}
\begin{table}[H]
\centering
\caption{Permutation importance comparison across models. Each feature’s mean importance and standard deviation are shown for HGB\_logRate, GBRT\_logRate, and RF\_logRate.}
\label{tab:r5}
\tiny
\setlength{\tabcolsep}{3pt}
\renewcommand{\arraystretch}{1.1}
\begin{tabularx}{\textwidth}{lrrrrrr}
\toprule
\multirow{2}{*}{\textbf{Feature}} 
& \multicolumn{2}{c}{\textbf{HGB}} 
& \multicolumn{2}{c}{\textbf{GBRT}} 
& \multicolumn{2}{c}{\textbf{RF}} \\
\cmidrule(lr){2-3} \cmidrule(lr){4-5} \cmidrule(lr){6-7}
& \textbf{Mean} & \textbf{Std} & \textbf{Mean} & \textbf{Std} & \textbf{Mean} & \textbf{Std} \\
\midrule
College Compelion               & 0.220 & 0.011 & 0.192 & 0.009 & 0.294 & 0.013 \\
Median Income                   & 0.183 & 0.010 & 0.199 & 0.010 & 0.184 & 0.008 \\
Stations Per 10k                & 0.076 & 0.007 & 0.080 & 0.008 & 0.059 & 0.006 \\
GDP                             & 0.053 & 0.003 & 0.062 & 0.003 & 0.043 & 0.003 \\
Total\_EVs\_2022                & 0.052 & 0.003 & 0.038 & 0.002 & 0.025 & 0.002 \\
Avg Gas Price                   & 0.050 & 0.004 & 0.047 & 0.005 & 0.036 & 0.002 \\
Charging Units Per 10k          & 0.022 & 0.003 & 0.020 & 0.002 & 0.053 & 0.005 \\
poverty                         & 0.019 & 0.001 & 0.013 & 0.001 & 0.008 & 0.001 \\
Avg Temp                        & 0.010 & 4.9$\times10^{-4}$ & 0.007 & 4.6$\times10^{-4}$ & 0.004 & 4.1$\times10^{-4}$ \\
Urbanicity                      & 0.008 & 7.5$\times10^{-4}$ & 0.006 & 7.8$\times10^{-4}$ & 0.009 & 1.0$\times10^{-3}$ \\
Sun Energy Gneration            & 0.007 & 8.2$\times10^{-4}$ & 0.007 & 1.3$\times10^{-3}$ & 0.003 & 2.7$\times10^{-4}$ \\
HS Graduate                     & 0.005 & 5.9$\times10^{-4}$ & 0.004 & 4.4$\times10^{-4}$ & 0.003 & 1.5$\times10^{-4}$ \\
Median AQI                      & 0.004 & 5.2$\times10^{-4}$ & 0.006 & 5.8$\times10^{-4}$ & 0.007 & 3.8$\times10^{-4}$ \\
Smoking Ban\_partial            & 9.6$\times10^{-4}$ & 1.7$\times10^{-4}$ & 6.1$\times10^{-4}$ & 2.1$\times10^{-4}$ & 2.6$\times10^{-4}$ & 2.3$\times10^{-5}$ \\
Smoking Ban\_comprehensive      & 5.6$\times10^{-4}$ & 1.5$\times10^{-4}$ & 0.0 & 0.0 & 6.1$\times10^{-6}$ & 2.0$\times10^{-6}$ \\
DEMOCRAT                        & 5.3$\times10^{-5}$ & 1.0$\times10^{-4}$ & 0.0 & 0.0 & 1.1$\times10^{-3}$ & 1.7$\times10^{-4}$ \\
REPUBLICAN                      & 4.4$\times10^{-6}$ & 7.2$\times10^{-5}$ & 5.4$\times10^{-5}$ & 4.8$\times10^{-5}$ & 8.3$\times10^{-4}$ & 1.2$\times10^{-4}$ \\
Smoking Ban\_none               & 0.0 & 0.0 & 0.0 & 0.0 & 1.4$\times10^{-4}$ & 1.2$\times10^{-5}$ \\
\bottomrule
\end{tabularx}
\end{table}

Table \ref{tab:r5} presents the mean importance of each feature and its standard deviation across the 25 runs in the cross validation for the top three models. It is evident that college completion and median income are consistently the most important predictors. The top three models assign similar importance weights with minimal variation, supporting the robustness of the predictive framework.

\begin{table}[H]
\centering
\caption{Average permutation importance across HGB\_logRate, GBRT\_logRate, and RF\_logRate models.}
\label{tab:r6}
\tiny
\setlength{\tabcolsep}{5pt}
\renewcommand{\arraystretch}{1.1}
\begin{tabularx}{0.5\textwidth}{l r}
\toprule
\textbf{Feature} & \textbf{Avg. Mean} \\
\midrule
College Compelion          & 0.235 \\
Median Income              & 0.189 \\
Stations Per 10k           & 0.072 \\
GDP                        & 0.053 \\
Total\_EVs\_2022           & 0.038 \\
Avg Gas Price              & 0.044 \\
Charging Units Per 10k     & 0.032 \\
poverty                    & 0.013 \\
Avg Temp                   & 0.007 \\
Urbanicity                 & 0.008 \\
Sun Energy Gneration       & 0.005 \\
HS Graduate                & 0.004 \\
Median AQI                 & 0.006 \\
Smoking Ban\_partial       & 6.1$\times10^{-4}$ \\
Smoking Ban\_comprehensive & 1.9$\times10^{-4}$ \\
DEMOCRAT                   & 4.8$\times10^{-4}$ \\
REPUBLICAN                 & 3.1$\times10^{-4}$ \\
Smoking Ban\_none          & 4.8$\times10^{-5}$ \\
\bottomrule
\end{tabularx}
\end{table}

Finally, we averaged each feature’s importance across the top three models , shown in Table \ref{tab:r6}. These averaged importance were subsequently used in the similarity network, where feature weights influence the computation of inter-county similarity. This makes the similarity graph model-aware, enabling us to examine whether counties with similar socio-politico-economic environmental profiles exhibit comparable levels of EV adoption.

\subsubsection{Model Comparisons with the Literature}

We compared our linear and nonlinear models with those reported in Kamis et al.~\cite{usa3}. There are several key methodological differences.
First, although the target variable in that study was also defined at the county level, all counties within a given state shared the same value, resulting in no within-state variation. This artificially increases model performance because a few features can easily differentiate states. In contrast, our dataset captures true county-level variation in EV adoption, producing a more realistic representation of socio-economic and political environmental effects.
Second, our target variable reflects EV registrations in 2023, whereas the previous study focused on 2021. Third, our features are more recent and temporally closer to the target (typically within one year). Fourth, our model selection is broader: we employ three linear models (Ridge, Elastic Net, Huber) and four nonlinear models (GBRT, HGB, RF, and MLP), while the previous study relied primarily on Multiple Linear Regression and Gradient Boosting. Notably, that study did not include the Histogram-Based Gradient Boosting or Random Forest models. Fifth, we evaluate performance using cross-validation and report metrics on the testing sets, whereas the previous study primarily reported the $R^2$ in the training performance.

Despite these differences, our linear models demonstrate superior performance compared with their linear baseline, achieving higher $R^2$ values. Specifically, the Ridge and Huber models achieve R2 scores of 0.649 and 0.627, respectively, compared to 0.621 in their Multiple Linear Regression model. For nonlinear models, our best model (GBRT) achieves $R^2=0.763$, while their best model (GBRT) achieves $R^2=0.98$ on the training set only. The methodological differences outlined above explain why our results, although lower in absolute value, are more reliable and generalizable to unseen data.

\begin{table}[H]
\centering
\caption{Cluster composition: number of counties per cluster and distribution by state.}
\label{tab:cluster_states}
\scriptsize
\setlength{\tabcolsep}{3pt}
\renewcommand{\arraystretch}{1.1}
\begin{tabularx}{\textwidth}{c c >{\raggedright\arraybackslash}X}
\toprule
\textbf{Cluster ID} & \textbf{\# of counties} & \textbf{state (counts)} \\
\midrule
9  & 33 & NJ (7), FL (4), VT (4), PA (3), VA (3), CO (2), IL (2), MD (2), TX (2), CT (1), MT (1), NY (1), OR (1) \\ \midrule
24 & 14 & CT (3), PA (3), FL (2), MD (2), NJ (2), IL (1), NC (1) \\ \midrule
13 & 33 & NJ (4), OH (4), CO (3), MD (3), MI (3), FL (2), IL (2), KY (2), TX (2), AL (1), IN (1), MN (1), NC (1), OR (1), PA (1), TN (1), WA (1) \\ \midrule
23 & 16 & FL (6), NC (3), VA (3), TX (2), AL (1), TN (1) \\ \midrule
26 & 13 & NY (7), CO (2), PA (2), NC (1), VA (1) \\ \midrule
2  & 48 & VA (13), IL (10), KY (5), IN (3), MI (3), AL (2), CO (2), NC (2), OH (2), PA (2), MD (1), NY (1), TN (1), TX (1) \\ \midrule
7  & 37 & TN (11), NC (9), VA (6), NY (4), MI (3), KY (1), OR (1), PA (1), WA (1) \\ \midrule
11 & 33 & OH (7), IL (6), TN (3), AL (2), CT (2), KY (2), LA (2), NC (2), PA (2), WI (2), FL (1), IN (1), MI (1) \\ \midrule
8  & 36 & IN (11), OH (7), KY (4), TN (3), MN (2), NY (2), PA (2), VA (2), MD (1), MI (1), NC (1) \\ \midrule
20 & 20 & WI (11), MN (5), MT (2), MI (1), UT (1) \\ \midrule
6  & 38 & MN (14), WI (13), MI (4), ME (2), NY (2), IN (1), OH (1), PA (1) \\ \midrule
12 & 33 & PA (10), OH (7), WI (7), MI (4), IL (1), KY (1), MN (1), UT (1), VT (1) \\ \midrule
22 & 18 & TX (17), LA (1) \\ \midrule
3  & 42 & TN (15), IN (10), KY (6), AL (5), NC (2), VA (2), NM (1), OH (1) \\ \midrule
10 & 33 & MN (9), IN (8), OH (8), KY (2), WI (2), IL (1), MD (1), MI (1), TN (1) \\ \midrule
27 & 9  & NC (3), TN (3), KY (2), NM (1) \\ \midrule
4  & 39 & OH (12), PA (6), IN (5), KY (5), IL (4), MI (4), MD (1), OR (1), WI (1) \\ \midrule
1  & 48 & MN (21), WI (9), MI (7), MT (3), OH (2), WA (2), IN (1), NY (1), PA (1), VA (1) \\ \midrule
25 & 14 & IL (12), CO (1), PA (1) \\ \midrule
16 & 26 & MI (11), MT (6), MN (2), PA (2), IL (1), ME (1), TN (1), UT (1), WI (1) \\ \midrule
19 & 21 & TX (15), IN (2), IL (1), KY (1), NC (1), VA (1) \\ \midrule
18 & 24 & KY (9), TN (7), AL (6), NC (1), NM (1) \\ \midrule
15 & 29 & TN (13), MI (5), NC (3), TX (3), MT (2), VA (2), LA (1) \\ \midrule
14 & 32 & KY (9), TX (8), LA (6), NC (3), TN (2), AL (1), MI (1), NM (1), OH (1) \\ \midrule
5  & 39 & IL (18), IN (12), OH (4), KY (3), MI (1), VA (1) \\ \midrule
21 & 18 & AL (5), TN (5), LA (4), KY (3), TX (1) \\ \midrule
17 & 26 & TX (23), FL (3) \\
\bottomrule
\end{tabularx}
\end{table}

\begin{figure}[h!]
    \centering
    \fbox{\includegraphics[width=\textwidth]{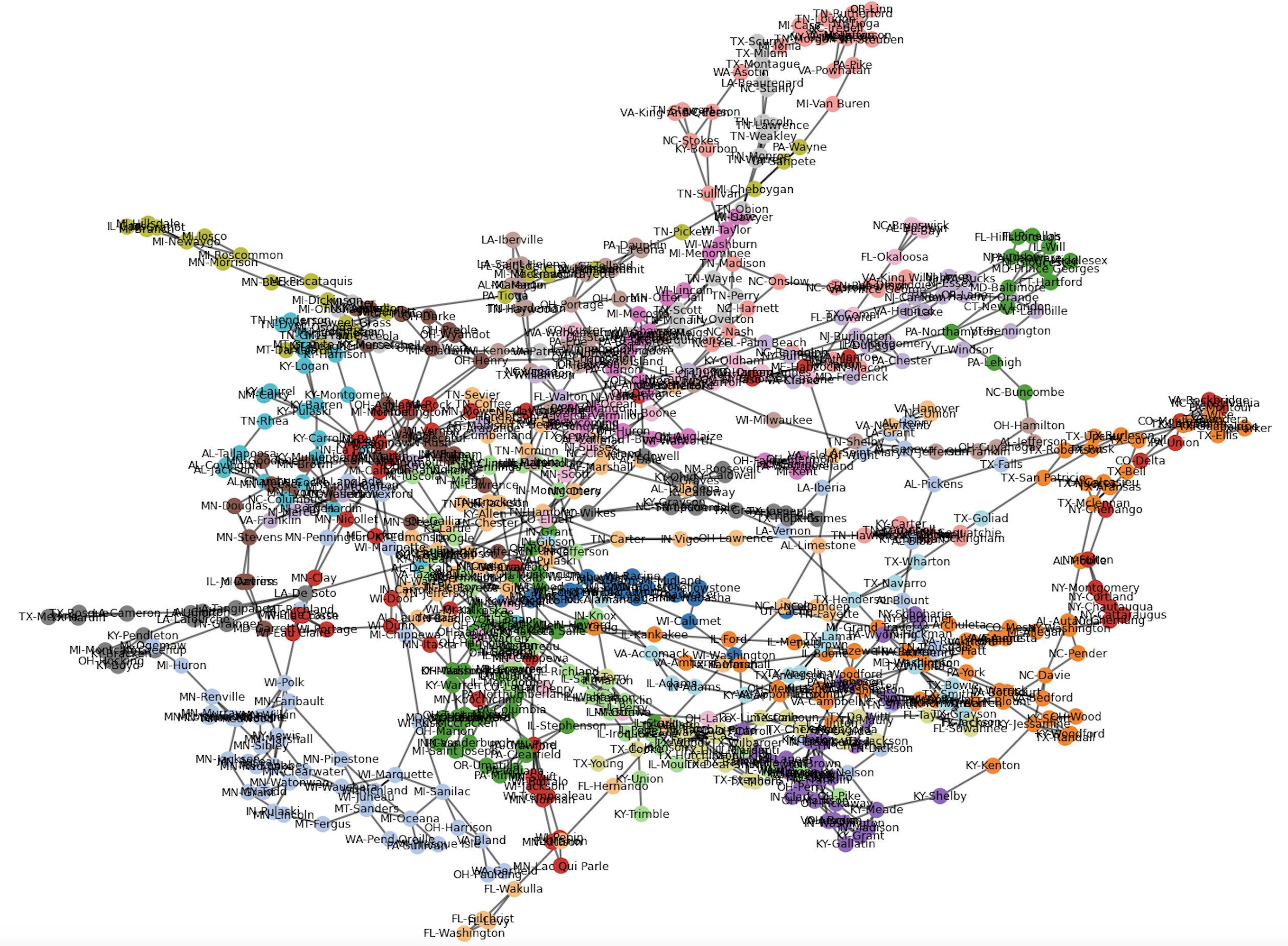}}
    \caption{County-level similarity network based on socio-economic, political, environmental, and infrastructure features. Each node represents a county, and edges denote high similarity between counties across the standardized feature space.}
    \label{fig:r2}
\end{figure}

\subsection{Theocratic Graph Based Analysis}

In this section, we present the results of the similarity network and the contextual structure revealed by the clusters. We then evaluate EV adoption across these clusters and identify the characteristics that most clearly differentiate them. Finally, we highlight the global feature patterns and the local variations that emerge within and between clusters.

\subsubsection{Similarity Network}

Fig \ref{fig:r2} shows the county-level similarity network, where nodes represent counties and edges represent strong similarities between them. We found that 752 out of 1,888 counties exhibit significant similarity in their socio-politico-economic, environmental, and infrastructural factors, which were weighted according to their importance in the predictive models.

The final similarity network selected by the optimal-$K$ procedure has an average degree of approximately three, which falls squarely within the stable connectivity range encouraged by the average-degree penalty in our methodology. This indicates that each county is, on average, directly connected to three of its most similar neighbors in the feature space. The resulting graph contains 1,173 edges, yielding an edge density of roughly 0.004.

Such sparsity is intentional; the mutual 
$k$-nearest-neighbor design retains only the strongest local similarities, avoiding the overly dense structures that the penalty function discourages. As a result, the network preserves interpretable local connectivity patterns that reveal meaningful community structure rather than uniform similarity across all counties.

Table~\ref{tab:cluster_states} presents the 27 clusters identified in the similarity network, along with the number of counties in each cluster, the states represented, and the distribution of counties across those states. A clear pattern of regional homogeneity emerges: many clusters are dominated by counties from only two or three states (e.g., three states dominate Cluster 2, while two states dominate Cluster 6). This indicates that counties grouped into the same cluster tend to share similar socio-economic, political, environmental, and infrastructural characteristics; those same characteristics weighted by their importance in the predictive models.

\begin{table}[H]
\centering
\caption{Summary of the global PERMANOVA test evaluating differences among the 27 EV adoption clusters.}
\label{tab:permanova_summary}
\scriptsize
\setlength{\tabcolsep}{8pt} 
\renewcommand{\arraystretch}{1.2} 
\begin{tabularx}{\textwidth}{l c c c c}
\toprule
\textbf{Test Method} & \textbf{Sample Size} & \textbf{Number of Groups} & \textbf{Test Statistic} & \textbf{p-value} \\
\midrule
PERMANOVA & 772 & 27 & 181.58 & 0.001 \\
\bottomrule
\end{tabularx}
\end{table}

    \begin{figure}[h!]
        \centering
        \includegraphics[width=0.8\textwidth]{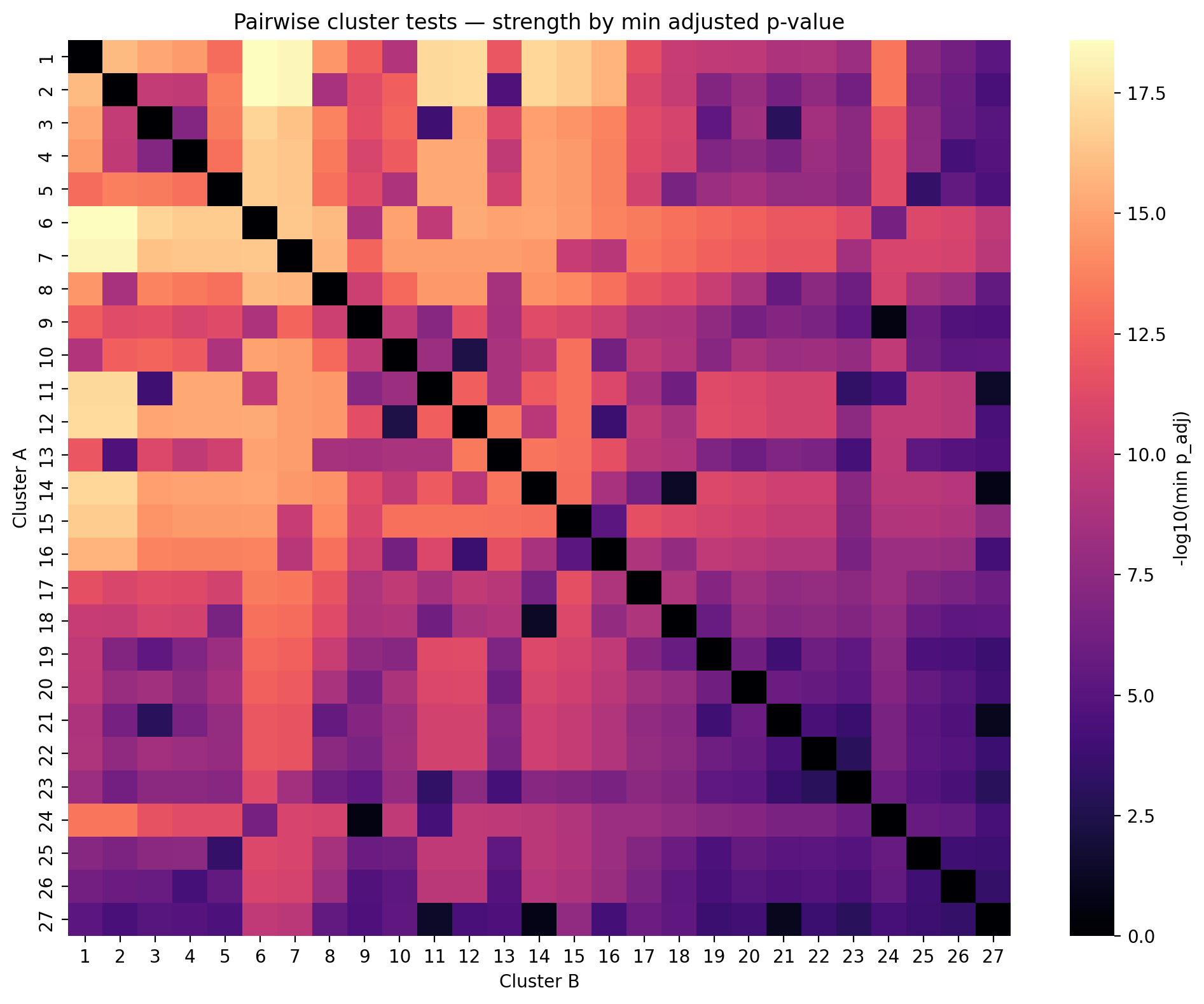}
        \caption{Pairwise cluster comparison heatmap showing the number of significantly different features between clusters. The color intensity represents the log-transformed p-values for better visualization of significance levels.}
        \label{fig:pairwise}
    \end{figure}

\subsubsection{Statistical Validation of Cluster Structure}

To statistically validate the cluster structure, we conducted a PERMANOVA test to evaluate whether all clusters are indistinguishable in terms of their weighted features. Table~\ref{tab:permanova_summary} summarizes the results. The null hypothesis, that no significant differences exist among clusters, was rejected, supported by an extremely small p-value (0.001).

We then performed pairwise PERMANOVA tests to determine which specific clusters differ from one another. A total of 351 pairwise comparisons were conducted across the 27 clusters, of which only four were not statistically significant. The remaining comparisons revealed clear distinctions among clusters. Figure~\ref{fig:pairwise} displays a heatmap of these pairwise results, in which p-values are shown using their negative logarithmic transformation to enhance visual interpretability.

\subsubsection{EV Adoption in Clusters}

\begin{figure}[h!]
    \centering
    \includegraphics[width=\textwidth]{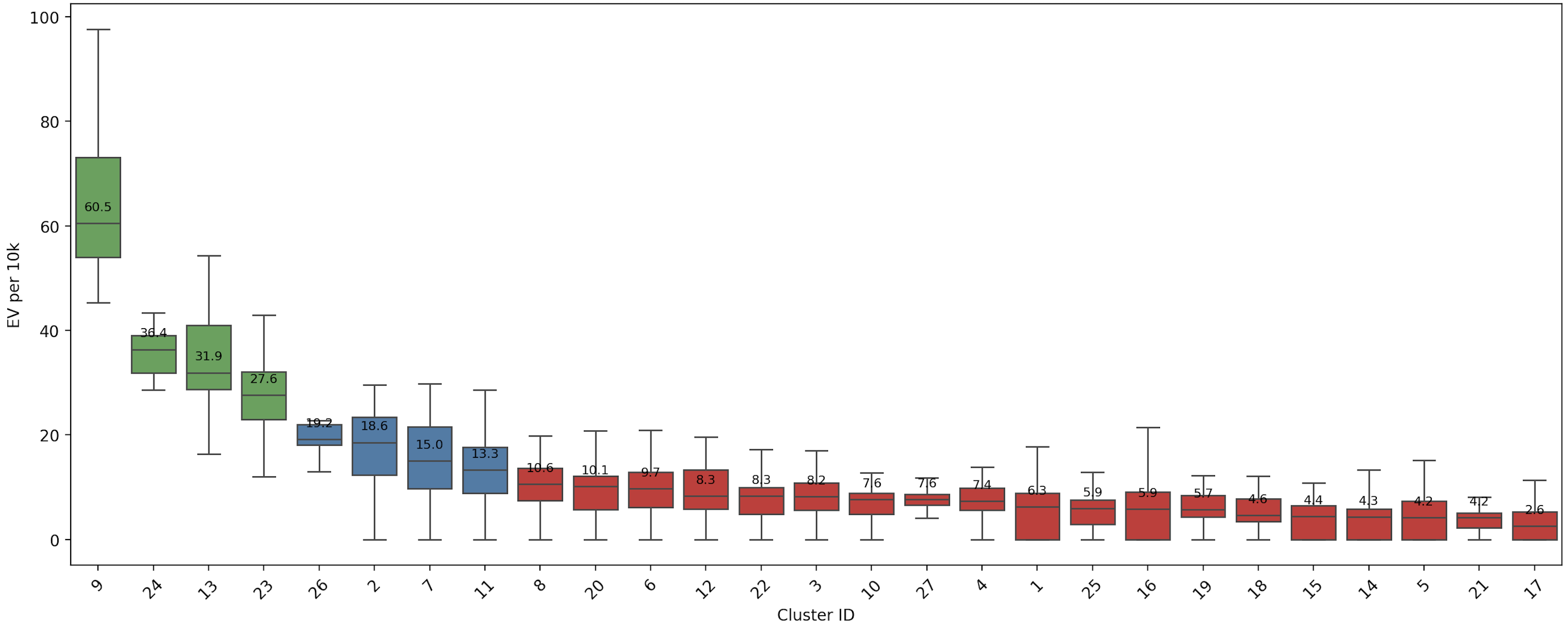}
    \caption{Boxplots of EV adoption (EVs per 10,000 people) across the 27 clusters. Each box represents the distribution of EV adoption within a cluster.}
    \label{fig:boxplot}
\end{figure}

The EV adoption rate as a target defined in Section~\ref{target}, is not included in the similarity computations between counties. Therefore, it serves as an external parameter that must be analyzed and compared across clusters. The enrichment analysis of the clusters, and the examination of how they characterize the EV adoption across counties, constitute a key component of this study. Figure \ref{fig:boxplot} presents boxplots for the 27 clusters, where each box depicts the distribution of EV adoption within a given cluster. Most of these distributions appear approximately normal, although the interquartile ranges and whisker lengths vary among clusters.

Consequently, the clusters were categorized into three EV adoption tiers of HAT, MAT, and LAT, based on their median values of the EV adoption. Clusters with medians in the range [0, 13] were assigned to the LAT, those in the range [13, 27] to the MAT, and those in the range [60, 100] to the HAT.

For each cluster, we constructed a socio-politico-economic, environmental, and infrastructural profile using the four most significant distinguishing features, determined by effect sizes (Cohen’s d), representing the standardized difference between the cluster mean and the mean of the remaining clusters for each feature. These features characterize the defining contextual attributes of each cluster, revealing distinct pathways through which different regions achieve varying levels of EV adoption.


\begin{table}[H]
\centering

\caption{Most significant features for the High Adoption Tier (HAT) clusters and their corresponding Cohen’s $d$ values.
Feature abbreviations: 
\textbf{CC} = College Completion; 
\textbf{PA(D)} = Political Affiliation (Democrat); 
\textbf{PA(R)} = Political Affiliation (Republican); 
\textbf{MI} = Median Income; 
\textbf{GDP} = Gross Domestic Product; 
\textbf{URB} = Urbanicity; 
\textbf{POV} = Poverty; 
\textbf{TEMP} = Average Temperature.}
\label{tab:HAT_features}
\resizebox{\textwidth}{!}{%
\begin{tabular}{|c|cc|cc|cc|cc|c|}
\hline
\multirow{2}{*}{Cluster id} &
\multicolumn{2}{c|}{Feature 1} &
\multicolumn{2}{c|}{Feature 2} &
\multicolumn{2}{c|}{Feature 3} &
\multicolumn{2}{c|}{Feature 4} &
\multirow{2}{*}{Tier} \\
\cline{2-9}
 & Feature & Cohen's $d$
 & Feature & Cohen's $d$
 & Feature & Cohen's $d$
 & Feature & Cohen's $d$ & \\
\hline
9  & CC    & 2.66 & PA(D) & 1.68 & MI   & 1.62 & GDP  & 1.46 & HAT \\
24 & PA(D) & 3.71 & CC    & 2.32 & URB  & -1.60& GDP  & 1.56 & HAT \\
13 & MI    & 1.65 & POV   & -1.60& CC   & 1.58 & URB  & -1.05& HAT \\
23 & TEMP  & 1.72 & URB   & -1.50& MI   & 0.63 & PA(R)& 0.58 & HAT \\
\hline
\end{tabular}%
}
\end{table}

\paragraph{\textbf{The High Adoption Tier (HAT)}}
This tier consists of four clusters, all exhibiting high levels of EV adoption compared with the other clusters found in the similarity network. Those clusters in HAT still have different median EV adoption rates. Therefore, each cluster can be characterized by the four most significant features in the similarity network, along with their corresponding Cohen’s d. The clusters in this tier are Clusters 9, 24, 13, and 23 with medians of 60.5, 36.4, 31.9, and 27.6, respectively as shown in Fig~\ref{fig:boxplot}. Table~\ref{tab:HAT_features} summarizes the key features of this tier clusters.

Cluster 9 includes 33 counties from different states as shown in Table~\ref{tab:cluster_states}. The strongest feature is College Completion, with a large effect size (Cohen's $d \approx 2.66$), indicating that the mean college completion rate in this cluster is substantially higher than in the rest. The second significant feature is Political Affiliation (Democrat) with Cohen’s d = 1.68, meaning most counties in this cluster lean Democratic. The third and fourth features are Median Income (d = 1.62) and GDP (d = 1.46), both suggesting that the counties in this cluster tend to be wealthier and have stronger economies than the others.

Cluster 24 contains 14 counties from different states as shown in Table~\ref{tab:cluster_states}. The most significant feature is Political Affiliation (Democrat) with Cohen’s d = 3.71. The second is College Completion (d = 2.32). The third is Urbanicity, which has a negative Cohen’s d (–1.60), suggesting that these counties are more urbanized than the others. The fourth feature is GDP (d = 1.56), indicating that these counties tend to have larger economies.

Cluster 13 includes 33 counties from different states as shown in Table~\ref{tab:cluster_states}. The first is Median Income (d = 1.65), suggesting it is a wealthy cluster. The second is Poverty, with a negative Cohen’s d (–1.60), indicating that this cluster has a lower poverty rate than the rest. The third feature is College Completion (d = 1.58), showing that this cluster is also well-educated. The fourth feature is Urbanicity (d = –1.05), meaning these counties are relatively more urban.

Cluster 23 contains 16 counties from different states as shown in Table~\ref{tab:cluster_states}. The most significant is Average Temperature (d = 1.72), indicating that these counties tend to be much warmer than the rest. The second is Urbanicity (d = –1.50), showing that this cluster is more urban. The third feature is Median Income with a moderate Cohen’s d (0.63), suggesting that average income in this cluster is higher than in the rest. The fourth feature is Political Affiliation (Republican) with Cohen’s d = 0.58.

Across all clusters in this tier, common characteristics include higher wealth, higher education, larger economies, and greater urbanization. However, the standardized differences of these features vary from cluster to cluster, distinguishing them from one another.

\begin{table}[H]
\centering
\caption{Most significant features for the Medium Adoption Tier (MAT) clusters and their corresponding Cohen’s $d$ values.
Feature abbreviations: 
\textbf{AGP} = Average Gas Price; 
\textbf{TEMP} = Average Temperature; 
\textbf{SB(N)} = Smoking Ban (None); 
\textbf{SB(P)} = Smoking Ban (Partial); 
\textbf{URB} = Urbanicity; 
\textbf{PA(D)} = Political Affiliation (Democrat); 
\textbf{PA(R)} = Political Affiliation (Republican); 
\textbf{MI} = Median Income; 
\textbf{POV} = Poverty; 
\textbf{CC} = College Completion; 
\textbf{AQI} = Median AQI; 
\textbf{GDP} = Gross Domestic Product; 
\textbf{ST10k} = Stations Per 10k.}
\label{tab:MAT_features}
\resizebox{\textwidth}{!}{%
\begin{tabular}{|c|cc|cc|cc|cc|c|}
\hline
\multirow{2}{*}{Cluster id} &
\multicolumn{2}{c|}{Feature 1} &
\multicolumn{2}{c|}{Feature 2} &
\multicolumn{2}{c|}{Feature 3} &
\multicolumn{2}{c|}{Feature 4} &
\multirow{2}{*}{Tier} \\
\cline{2-9}
 & Feature & Cohen's $d$
 & Feature & Cohen's $d$
 & Feature & Cohen's $d$
 & Feature & Cohen's $d$ & \\
\hline
26 & AGP & 1.20 & TEMP & -0.92 & SB(N) & 0.76 & ST10k & 0.70 & MAT \\
2  & URB & -1.06 & POV & -0.95 & MI & 0.69 & CC & 0.65 & MAT \\
7  & SB(P) & 4.46 & SB(N) & -3.00 & URB & -1.00 & PA(R) & 0.59 & MAT \\
11 & PA(D) & 4.14 & AQI & 0.87 & URB & -0.79 & GDP & 0.78 & MAT \\
\hline
\end{tabular}%
}
\end{table}

\paragraph{\textbf{The Medium Adoption Tier (MAT)}}
This tier ranks second in terms of EV adoption. Clusters 26, 2, 7, and 11 fall into this tier are  as shown in Fig~\ref{fig:boxplot}, with medians of 19.2, 18.6, 15.0, and 13.3, respectively. Table~\ref{tab:MAT_features} summarizes the key features of this tier clusters.

Cluster 26 includes 16 counties as shown in Table~\ref{tab:cluster_states}. The most significant feature is Average Gas Price (d = 1.20), suggesting that gas prices in this cluster are much higher than in the rest. The second feature is Smoking Ban (None) (d = 0.76), indicating a lack of smoking restrictions. The third feature is Average Temperature (d = –0.92), showing that these counties tend to be colder. The fourth feature is Stations per 10000 people (d= 0.7) reflecting that this cluster has strong charging infrastructure than the rest. 

Cluster 2 includes 48 counties as shown in Table~\ref{tab:cluster_states}. The first is Urbanicity (d = –1.06), suggesting that these counties are more urbanized. The second is Poverty (d = –0.95), indicating lower poverty rates. The third is Median Income (d = 0.69), showing that this cluster is relatively wealthier. The fourth is College Completion (d = 0.65), indicating higher education levels than the rest.

Cluster 7 includes 37 counties as shown in Table~\ref{tab:cluster_states}. The first and second features are smoking-related policies: Partial Smoking Ban (d = 4.46) and No Smoking Ban (d = –3.00). The third feature is Urbanicity (d = –1.00), and the fourth is Political Affiliation (Republican) (d = 0.59). Overall, this cluster appears to be policy-oriented and more conservative.

Cluster 11 includes 33 counties as shown in Table~\ref{tab:cluster_states}. The first is Political Affiliation (Democrat) with a large effect size (d = 4.14), indicating that nearly all counties in this cluster lean Democratic. The second feature is Median AQI (d = 0.87), meaning air quality is poorer than the average of the rest. The third is Urbanicity (d = –0.79), suggesting greater urbanization. The fourth is GDP (d = 0.78), indicating more industrialized and economically strong counties.

Clusters in the MAT show less homogeneity in their significant features compared with those in the HAT. However, urbanicity appears to be a primary driver of medium-level EV adoption, followed by wealth-related features such as GDP and median income. The clusters within this tier can be differentiated by several features and are clearly distinct from those in the HAT.

\begin{table}[H]
\centering
\caption{Most significant features for the Low Adoption Tier (LAT) clusters and their corresponding Cohen’s $d$ values.
Feature abbreviations: 
\textbf{URB} = Urbanicity; 
\textbf{SB(C)} = Smoking Ban (Comprehensive);
\textbf{SB(N)} = Smoking Ban (None); 
\textbf{POV} = Poverty; 
\textbf{PA(R)} = Political Affiliation (Republican); 
\textbf{TEMP} = Average Temperature; 
\textbf{GAS} = Average Gas Price; 
\textbf{HSG} = High School Graduation; 
\textbf{CC} = College Completion; 
\textbf{AQI} = Median AQI.
\textbf{MI} = Median Income;}

\label{tab:LAT_features}
\resizebox{\textwidth}{!}{%
\begin{tabular}{|c|cc|cc|cc|cc|c|}
\hline
\multirow{2}{*}{Cluster id} &
\multicolumn{2}{c|}{Feature 1} &
\multicolumn{2}{c|}{Feature 2} &
\multicolumn{2}{c|}{Feature 3} &
\multicolumn{2}{c|}{Feature 4} &
\multirow{2}{*}{Tier} \\
\cline{2-9}
 & Feature & Cohen's $d$
 & Feature & Cohen's $d$
 & Feature & Cohen's $d$
 & Feature & Cohen's $d$ & \\
\hline
8  & URB & -1.79 & SB(N) & 0.77 & POV & -0.68 & PA(R) & 0.59 & LAT \\
20 & TEMP & -1.64 & HSG & 1.33 & POV & -1.29 & URB & -0.97 & LAT \\
21 & HSG & -1.43 & TEMP & 1.30 & CC & -1.11 & POV & 0.98 & LAT \\
17 & TEMP & 1.88 & HSG & -1.70 & POV & 1.27 & AQI & 1.27 & LAT \\
6 & PA(D) & 4.29 & TEMP & -1.58 & AQI & -1.48 & HSG & 1.02 & LAT \\

12 & SB(C) & 6.54 & SB(N) & -2.96 & TEMP & -1.02 & HSG & 0.72 & LAT \\

22 & TEMP & 2.52 & GAS & -1.6 & URB & -1.3p & AQI & 1.19 & LAT \\

3 & MI & -0.88 & CC & -0.70 & SB(N) & 0.65 & POV & 0.62 & LAT \\

10 & URB & 1.12 & POV & -0.96 & HSG & 0.83 & TEMP & -0.75 &  LAT \\

27 & HSG & -1.84 & MI & -1.42 & POV & 1.26 & GAS & -1.14 & LAT \\

4 & SB(N) & 0.78 & PA(D & 0.59 & SB(P) & -0.57 & MI & -0.49 & LAT \\

1 & URB & 2.25 & TEMP & -1.75 & AQI & -1.31 & CC & -0.64 & LAT \\

25 & AQI & 1.66 & GAS & 1.55 & URB & 1.11 & SB(N) & 0.76 & LAT \\

16 & SB(P) & 4.14 & SB(N) & -2.89 & URB & 1.68 & TEMP & -1.58 & LAT \\
19 & GAS & -1.15 & TEMP & 1.41 & CC & -0.82 & SB(N) & 0.76 & LAT \\
18 & POV & 1.55 & HSG & -1.51 & MI & -1.49 & CC & -1.05 & LAT \\

15 & SB(P) & 4.22 & SB(N) & -2.92 & URB & 1.81 & MI & -1.16 & LAT \\

14 & SB(C) & 6.44 & SB(N) & -2.95 & HSG & -1.18 & POV & 1.11 & LAT \\

5 & URB & 1.63 & CC & -1.14 & SB(N) & 0.78  &  MI & -0.74 & LAT \\

\hline
\end{tabular}%
}
\end{table}

\paragraph{\textbf{The Low Adoption Tier (LAT)}}
Most clusters, 19 out of the 27 clusters, fall within this tier. The EV adoption decreases as we move away from the HAT and MAT, which is evident from the trajectory of the medians in Fig~\ref{fig:boxplot}. In this tier, we discuss two clusters adjacent to the MAT and two from the far right. Table~\ref{tab:LAT_features} summarizes the key features of this tier clusters.

Clusters 8 and 20 represent the highest EV adoption levels within the LAT, containing 36 and 20 counties, respectively, as shown in Table~\ref{tab:cluster_states}. For Cluster 8, the most significant feature is Urbanicity (d = –1.79), suggesting these counties are relatively urbanized, similar to some clusters in higher tiers. The second is Smoking Ban (None) (d = 0.77), indicating a relative lack of smoking restrictions. The third is Poverty (d = –0.68), showing lower poverty rates. The fourth is Political Affiliation (Republican) (d = 0.59). For Cluster 20, the most significant feature is Average Temperature (d = –1.64), suggesting that these counties are colder than the rest. The second is High School Graduation (d = 1.33), indicating higher graduation rates. The third is Poverty (d = –1.29), showing lower poverty, and the fourth is Urbanicity (d = –0.97), suggesting more urbanized environments.

Clusters 21 and 17 fall at the lower end of the LAT, with 18 and 26 counties and medians of 4.2 and 2.6 EVs per 10,000 people, respectively. For Cluster 21, the four most significant features are High School Graduation (d = –1.43), Average Temperature (d = 1.30), College Completion (d = –1.11), and Poverty (d = 0.98). These results indicate low education levels, higher poverty, and warmer climates. For Cluster 17, the main features are Average Temperature (d = 1.88), High School Graduation (d = –1.70), Poverty (d = 1.27), and Median AQI (d = 1.27). This cluster shares three features with Cluster 21 but differs in magnitude: Cluster 17 has lower high school graduation rates, slightly cooler temperatures, and higher poverty.

This tier contains many clusters, each identifiable by its four most significant features and corresponding Cohen’s d. As we move leftward toward the MAT, clusters become more similar to those in higher tiers—being more educated, wealthier, and urbanized. In contrast, moving rightward within this tier reveals clusters that are less educated, poorer, and located in warmer regions.

\subsubsection{Structural Trends and Contextual Deviations in EV Adoption Drivers}

\begin{figure}[h!]
    \centering
    \includegraphics[width=\textwidth]{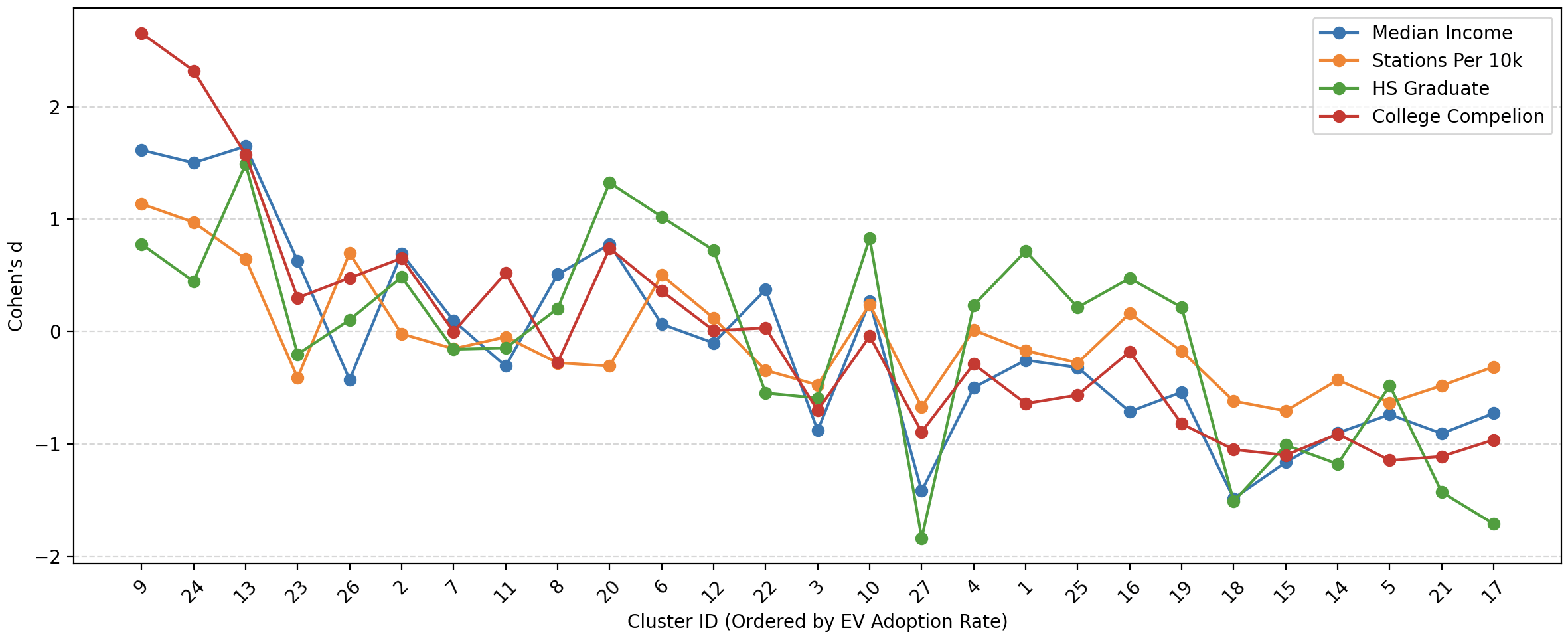}
    \caption{Standardized mean differences (Cohen’s d) for four negatively trending features—median income, charging stations per 10,000 population, high school graduation rate, and college completion rate—across clusters ordered by EV adoption rate. These structural features generally decline as EV adoption decreases, with minor variations between adjacent clusters.}
    \label{fig:negative}
\end{figure}

Fig~\ref{fig:negative} shows that across the 27 clusters, the selected socioeconomic and infrastructural features (median income, charging stations per 10,000 population, high school graduation rate, and college completion rate) generally follow clear global trends that align with the EV adoption gradient. Such features exhibit a consistent negative trend: effect sizes decrease steadily as we move from high- to low-adoption clusters. Although minor local fluctuations exist (where a low-adoption cluster may temporarily show a slightly higher value than the cluster preceding it) the overall direction remains monotonic. Importantly, these four structural features show strong internal agreement within each cluster, confirming that they function as broad, system-level determinants of EV adoption across the region.

\begin{figure}[h!]
    \centering
    \includegraphics[width=\textwidth]{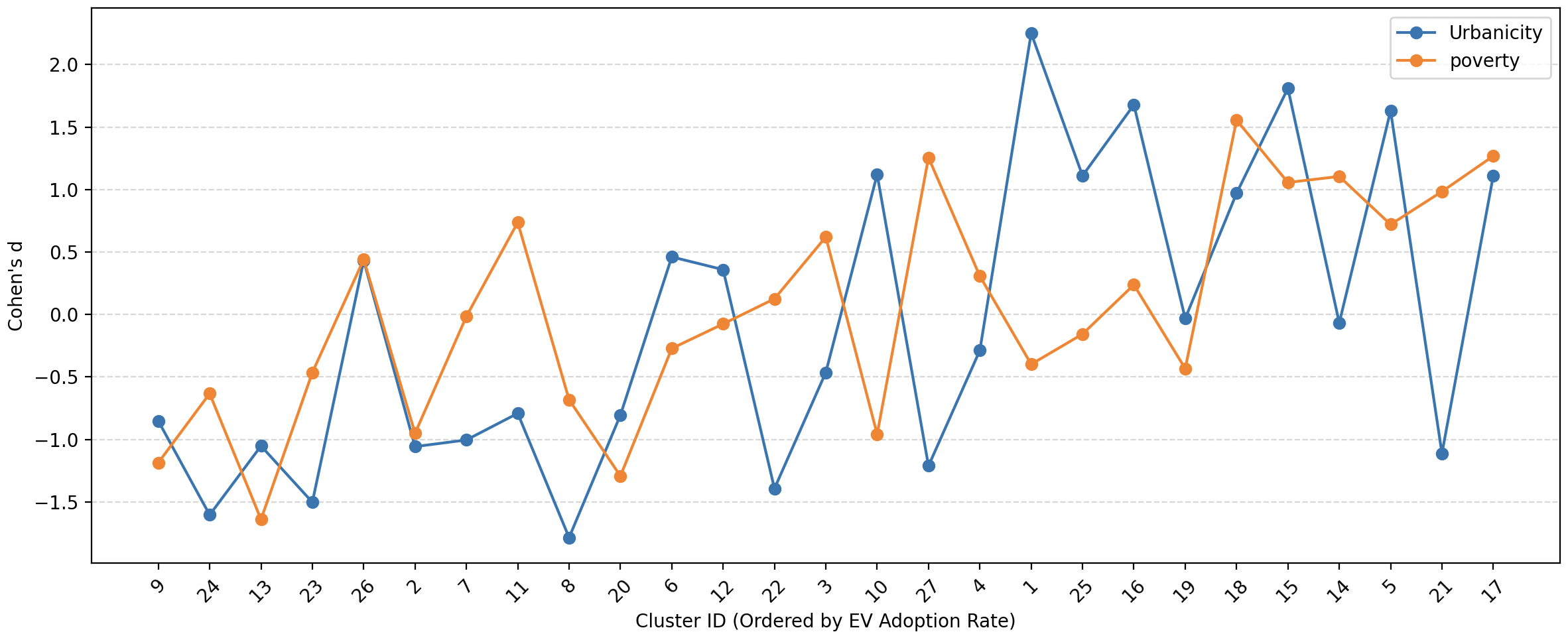}
    \caption{Standardized mean differences (Cohen’s d) for urbanicity and poverty rate across clusters ordered by EV adoption rate. Both features exhibit an overall positive trend moving toward lower-adoption clusters, with notable variations among clusters within the Low Adoption Tier.}
    \label{fig:positive}
\end{figure}

In contrast, Fig~\ref{fig:positive} show that the two positively trending features (urbanicity and poverty) reveal a more complex pattern that departs from the uniformity seen in the structural variables. While both features increase in magnitude as adoption decreases, they display notable disagreements within the LAT. Specifically, Clusters 1, 25, and 16 exhibit low urbanicity but also low poverty, indicating that rurality, rather than economic disadvantage, might be the primary barrier to EV adoption in these regions. Conversely, Clusters 27, 14, and 21 show high urbanicity alongside high poverty rates, revealing that economic constraints, rather than urban form, might be the dominant barrier in these urban but financially constrained areas.

These subgroup differences are further supported by the cluster-characterization results in Table~\ref{tab:LAT_features}: poverty emerges as a significant feature for the urbanized–high-poverty clusters (27, 14, 21), while urbanicity is a key distinguishing feature for the rural–low-poverty clusters (1, 25, 16). Thus, although all six clusters fall within the same tier, the underlying causes of low EV uptake differ substantially. This divergence shows that structural features collectively shape the global adoption gradient, whereas contextual features differentiate the local mechanisms driving low adoption within the same tier.

These findings highlight a central strength of the graph-theoretic clustering approach. Unlike traditional trend-based analysis or predictive modeling, which tend to capture only global effects and smooth out local deviations, the graph method identifies distinct pathways leading to similar adoption outcomes. By uncovering cluster-specific combinations of rurality, poverty, infrastructure gaps, and socioeconomic profiles, this approach provides authorities with actionable, localized insights. Such fine-grained differentiation is essential for designing targeted strategies, allowing policymakers to tailor EV promotion programs to the unique barriers present in each cluster, rather than relying on one-size-fits-all interventions.

\section{Conclusion}
\label{conclusion}

This study introduced a graph-theoretic framework that complements prediction-based analyses to provide a deeper and more nuanced understanding of the factors shaping EV adoption across U.S. counties. By integrating weighted feature importances from multiple predictive models into a weighted Gower similarity metric, we constructed a county similarity network and identified 27 clusters using modularity-based community detection. A global PERMANOVA test, supported by nearly all pairwise comparisons, confirmed that these clusters differ meaningfully in their combined socioeconomic, political, environmental, and infrastructural characteristics.

Analysis of EV adoption across these clusters revealed that the clusters themselves naturally fall into three EV adoption tiers: High, Medium, and Low Adoption Tiers. Consistent structural trends (such as declining median income, educational attainment, and charging-station availability from higher to lower tiers) highlight broad national patterns affecting adoption. However, the cluster-level effect size analysis also uncovered important local deviations that global trend analysis or prediction models alone cannot capture. Specifically, while income, education, and station availability follow stable monotonic patterns, the positively trending features (such as urbanicity and poverty) show substantial disagreement within the Low Adoption Tier. Some low-adoption clusters are rural yet not economically disadvantaged, whereas others are urbanized but experience high poverty rates. These results demonstrate that different mechanisms can lead to the same low-adoption outcome, emphasizing that barriers to EV adoption vary significantly even within the same adoption tier.

By revealing both global structural gradients and localized barriers, the graph-theoretic approach provides policymakers with actionable, cluster-specific insights. Understanding whether low adoption arises from rural infrastructure constraints, economic hardship, or other contextual factors enables more effective, targeted strategies for accelerating EV uptake. Rather than relying on uniform nationwide interventions, decision-makers can design policies that respond to the unique socioeconomic and infrastructural characteristics of each cluster.
Although the dataset used in this study includes EV information for 1,888 of the 3,244 U.S. counties, limited data availability remains a challenge for EV-related research. Future work could incorporate additional counties as more comprehensive datasets become accessible. Another promising direction involves constructing temporal similarity or correlation networks using historical EV adoption time series, enabling the analysis of dynamic adoption trajectories and cluster evolution over time. Such extensions depend on expanded access to longitudinal county-level EV data, which would further strengthen evidence-based EV adoption planning.

\bibliographystyle{unsrt}  

\bibliography{references}

\end{document}